\documentclass[twocolumn,showpacs,prl]{revtex4}
\usepackage{graphicx}
\usepackage{color}
\usepackage{amssymb}
\newcommand{\xrm}[1]{{\textstyle \mbox{\rm #1}}}

\newcommand{\abs}[1]{\left| #1\right|}
\begin{document}
\title{Reply to Comment on ``Material evidence of a 38 MeV boson''}
\author{Eef van Beveren$^{1}$ and George Rupp$^{2}$}
\affiliation{
$^{1}$Centro de F\'{\i}sica Computacional,
Departamento de F\'{\i}sica, Universidade de Coimbra,
P-3004-516 Coimbra, Portugal\\
$^{2}$Centro de F\'{\i}sica das Interac\c{c}\~{o}es Fundamentais,
Instituto Superior T\'{e}cnico, Technical University of Lisbon,
P-1049-001 Lisboa, Portugal
}
\date{\today}

\begin{abstract}
We reply to a very recent Comment \cite{ARXIV12042349}
by Bernhard, Friedrich, Schl\"{u}ter, and Sch\"{o}nning
for the COMPASS Collaboration, in which it is stated
that Monte-Carlo simulations of COMPASS data do not support
any interpretation of a peaked structure in diphoton invariant-mass
distributions below the $\pi^{0}$ mass in terms of a new
resonance.  Here we show, by directly comparing the simulations
to the COMPASS data themselves, that the authors' claim is not
substantiated by the Monte-Carlo results presented in the
Comment.
\end{abstract}

\pacs{11.15.Ex,
%12.10.Dm,
12.38.Aw, 12.39.Mk, 14.80.Ec}

\maketitle

Before directly replying to the Comment \cite{ARXIV12042349}
of Bernhard, Friedrich, Schl\"{u}ter, and Sch\"{o}nning (BFSS)
(for the COMPASS Collaboration) on our interpretation
\cite{ARXIV12021739}of a resonance-like structure in recent
COMPASS data \cite{ARXIV11086191,ARXIV11090272},
let us first recapitulate some of the essential features of our work,
for those readers who are not so familiar with it.

\section{Motivation}

In Ref.~\cite{NCA80p401} an $SO(4,2)$ conformally symmetric model
was proposed for strong interactions at low energies,
based on the observation \cite{THEFNYM7911,PRD30p1103,LNP211p331}
that confinement can be described
by an anti-De Sitter (aDS) background geometry.
The possibility of such a strategy had already been studied,
almost a century ago, by H.~Weyl \cite{AdP364p101},
who found that the dynamical equations of gauge theories
retain their flat-space-time form
when subject to a conformally-flat metrical field
instead of the usual Minkowski background.
The unification of electromagnetism and strong interactions
can be justified by the very subtle balance between these forces
in the nucleus, where just one neutron more or less
can make the difference between stability or instability.

Confinement of quarks and gluons is modeled
by the introduction of two scalar fields which
spontaneously break the $SO(4,2)$ symmetry down to
$SO(3,2)$ and $SO(3)\otimes SO(2)$ symmetry, respectively.
Moreover, a symmetric second-order tensor field is defined
that serves as the metric for flat space-time,
coupling to electromagnetism.
Quarks and gluons, which to lowest order
do not couple to this tensor field,
are confined to an aDS universe \cite{ARXIV07061887},
having a finite radius in the flat space-time.
This way, the model describes quarks and gluons, which
oscillate with a universal frequency ---
independent of the flavor mass --- inside a closed universe,
as well as photons, which freely travel through flat space-time.

The fields in the model of Ref.~\cite{NCA80p401}
comprise one real scalar field $\sigma$
and one complex scalar field $\lambda$.
Their dynamical equations were solved in Ref.~\cite{NCA80p401}
for the case that the respective vacuum expectation values,
given by $\sigma_{0}$ and $\lambda_{0}$,
satisfy the relation
\begin{equation}
\abs{\sigma_{0}}\gg \abs{\lambda_{0}}
\;\;\; .
\label{slvacua}
\end{equation}
A solution for $\sigma_{0}$ of particular interest
leads to aDS confinement, via the associated
conformally flat metric given by $\sigma\eta_{\mu\nu}$.

The only quadratic term in the Lagrangian of Ref.~\cite{NCA80p401}
is proportional to
\begin{equation}
-\sigma^{2}\lambda^{\ast}\lambda
\;\;\; .
\label{quadraticterm}
\end{equation}
Hence, under the condition of relation (\ref{slvacua}),
one obtains, after choosing vacuum expectation values,
a light $\sigma$ field associated with confinement,
and a very heavy complex $\lambda$ field
associated with electromagnetism.
Here, we will study the --- supposedly light --- mass
of the scalar field that gives rise to confinement.

The conformally symmetric model of Ref.~\cite{NCA80p401}
in itself does not easily allow for interactions between hadrons,
as each hadron is described by a closed universe.
Therefore, in order to compare the properties of this model
to the actually measured cross sections and branching ratios,
the model has been further simplified,
such that only its main property survives,
namely its flavor-independent oscillations.
This way the full aDS spectrum is, via light-quark-pair creation,
coupled to the channels of two --- or more --- hadronic decay products
for which scattering amplitudes can be measured,
thus relating observed resonances to the spectrum of aDS.

The aDS spectrum reveals itself through the structures
observed in hadronic invariant-mass distributions.
However, as we have shown in the past
(see Ref.~\cite{ARXIV10112360} and references therein),
there exists no simple relation between
enhancements in the experimental cross sections
and the aDS spectrum.
Nevertheless, this was studied in parallel, for mesons,
in a coupled-channel model
in which quarks are confined by
a flavor-independent harmonic oscillator
\cite{PRD21p772,PRD27p1527}.
Empirically, based on numerous data on mesonic resonances
measured by a large variety of experimental collaborations,
it was found \cite{AIPCP1374p421}
that an aDS oscillation frequency of
\begin{equation}
\omega =190
\;\;\;\xrm{MeV}
\label{uniosc}
\end{equation}
agrees well with the observed results for
meson-meson scattering and meson-pair production
in the light \cite{ZPC30p615},
heavy-light \cite{PRL91p012003},
and heavy \cite{CNPC35p319,ARXIV10094097} flavor sectors,
thus reinforcing the strategy proposed in Ref.~\cite{NCA80p401}.

A further ingredient of the model for the description
of non-exotic quarkonia, namely the coupling
of quark-antiquark components
to real and virtual two-meson decay channels \cite{AP324p1620}
via $^{3\!}P_{0}$ quark-pair creation,
gives us a clue about the size of the mass of the $\sigma$ field.
For such a coupling it was found that the average radius $r_{0}$
for light-quark-pair creation in quarkonia can be described
by a flavor-independent mass scale, given by
\begin{equation}
M=\frac{1}{2}\omega^{2}\mu r_{0}^{2}
\;\;\; ,
\label{unirad}
\end{equation}
where $\mu$ is the effective reduced quarkonium mass.
In earlier work, the value $\rho_{0}=\sqrt{\mu\omega}r_{0}=0.56$
\cite{PRD21p772,PRD27p1527} was used,
which results in $M=30$ MeV for the corresponding mass scale.
However, the quarkonium spectrum is not very sensitive
to the precise value of the radius $r_{0}$,
in contrast with the resonance widths.
In more recent work \cite{HEPPH0201006,PLB641p265},
slightly larger transition radii have been applied,
corresponding to values around 40 MeV for $M$.
Nevertheless, values of 30--40 MeV
for the flavor-independent mass $M$
do not seem to bear any relation
to an observed quantity for strong interactions.
However, in Refs.~\cite{ARXIV11021863,ARXIV12021739}
we have presented experimental evidence
for the possible existence of a quantum with a mass of about 38 MeV,
which in light of its relation to the $^{3\!}P_{0}$ mechanism
we suppose to mediate quark-pair creation.
Moreover, its scalar properties make it a perfect candidate
for the quantum associated with
the above-discussed scalar field for confinement.

\section{First observations}

In recent papers \cite{ARXIV11021863,ARXIV12021739},
we have presented a variety of indications of the possible existence
of a light boson with a mass of about 38 MeV,
henceforth referred to as $E(38)$.
These indications amounted to a series of
low-statistics observations all pointing in the same direction.
Each of the results alone, viz.\
interference effect \cite{PRD79p111501R},
small flavor-independent oscillations
in electron-positron and proton-antiproton annihilation data
\cite{ARXIV10095191},
the excess signals visible \cite{ARXIV11021863,ARXIV12021739}
in the $\mu^{+}\mu^{-}$ mass distributions of
$\Upsilon\left( 2\,{}^{3\!}S_{1}\right)$ $\to$
$\pi^{+}\pi^{-}\Upsilon\left( 1\,{}^{3\!}S_{1}\right)$
$\to$ $\pi^{+}\pi^{-}\mu^{+}\mu^{-}$,
in $\Upsilon\left( 3\,{}^{3\!}S_{1}\right)$ $\to$
$\pi^{+}\pi^{-}\Upsilon\left( 1\,{}^{3\!}S_{1}\right)$
$\to$ $\pi^{+}\pi^{-}\mu^{+}\mu^{-}$,
in $\Upsilon\left( 3\,{}^{3\!}S_{1}\right)$ $\to$
$\pi^{+}\pi^{-}\Upsilon\left( 2\,{}^{3\!}S_{1}\right)$
$\to$ $\pi^{+}\pi^{-}\mu^{+}\mu^{-}$,
and in the $e^{+}e^{-}$ mass distributions of
$e^{+}e^{-}$ $\to$ $\pi^{+}\pi^{-}e^{+}e^{-}$,
and finally hybrid signal \cite{ARXIV11021863},
has much too low statistics to make firm claims.
However, we noted that all these results point in the same direction.

Indeed, it seems highly unlikely that by sheer coicidence one finds
the same kind of oscillations in four different sets of data
\cite{PRD79p111501R,ARXIV10095191} involving different flavors,
statistical fluctuations at $\pm 38$ MeV
in yet another four sets of different data,
and finally a resonance-like fluctuation at $38$ MeV
in a further set of data.
Moreover, the resulting scalar mass comes out where it was
predicted via our analyses in mesonic spectroscopy
(see Ref.~\cite{ARXIV10112360} and references therein).

\section{Enhancement around 40 MeV in COMPASS data}

\begin{figure}[htpb]
\begin{center}
\begin{tabular}{c}
\includegraphics[width=230pt]{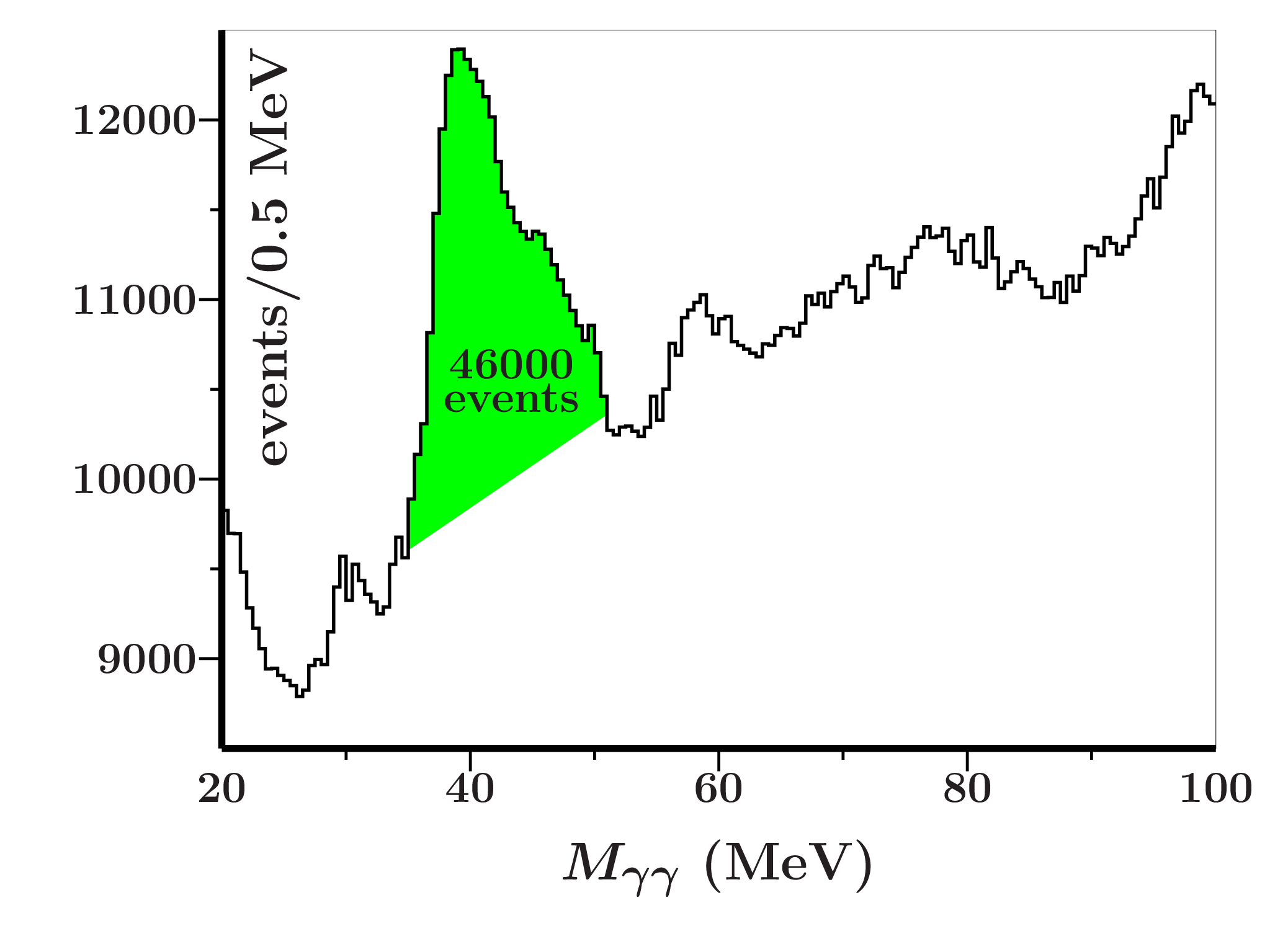}\\ [-20pt]
\end{tabular}
\end{center}
\caption{\small
Detail of the invariant two-photon mass distribution of COMPASS
\cite{ARXIV11086191}. We observe a clear signal, with maximum at
about 39 MeV.
}
\label{compassgammagamma}
\end{figure}

In Ref.~\cite{ARXIV12021739}, we presented more pieces
of evidence, one of which, viz.\ a small resonance-like signal in the
diphoton invariant-mass distribution \cite{ARXIV11086191} of the
COMPASS Collaboration, is considerably more conclusive
than previously reported \cite{ARXIV11021863} signals, owing to
much higher statistics.
The data, depicted in Fig.~\ref{compassgammagamma},
have been obtained at the two-stage
magnetic COMPASS spectrometer attached to the SPS accelerator
facility at CERN \cite{ARXIV11086191}.
These data seem to have enough statistics to substantiate the existence of
a light boson with mass around 40 MeV.
However, in a more recent version \cite{ARXIV11086191} of their work,
the COMPASS Collaboration added the following remark
to the figure caption regarding the enhancement in
Fig.~\ref{compassgammagamma}.
\begin{quote}
\em
``The structures below the $\pi^{0}$ mass peak are artefacts
of low energetic photon reconstruction due to secondary interactions
in the detector material and to cuts in the reconstruction algorithm.
They should not be mistaken for any physical signal.''
\em
\end{quote}

Although it may of course be possible to obtain resonance-like structures
by artefacts in data collection, we are convinced this is not the case
for the signal in the 40 MeV region, because it coincides
surprisingly well with the other observations reported
in Refs.~\cite{ARXIV11021863,ARXIV12021739}.

Furthermore, it is clear that the light boson cannot be
an ordinary hadron emerging from a hadronic vertex,
unless at an extremely low rate.
Otherwise, it would have been observed long ago.
Ordinary hadronization in high-energy collisions
gives rise to pions, kaons, and other hadrons that are stable with
respect to strong interactions.
These are processes in which quark-pair creation and gluonic jets dominate.
But on the other hand, judging from the amount of events in the
low-mass enhancement in the two-photon data,
which is about 10\% of the number of events in the $\eta$ signal,
it does seem to be produced with a reasonably large rate
in the COMPASS experiment.
Such a rate indicates that it most probably is a hadronic particle,
though with very peculiar properties that still have to be
understood.

In this paper we shall assume that the $E(38)$ boson has the shape
of a spherical bubble, as predicted by anti-De Sitter confinement
\cite{THEFNYM7911,JMP27p1411}, i.e., a thin film of glue.
Due to possible surface oscillations,
such a system almost has a continuum of excitations.
If we assume that the stability of the bubble rapidly decreases
for excitations and allow for a spreading of some 20 MeV,
then we may fully reconstruct the line shape
shown in Fig.~\ref{compassgammagamma}.
Moreover, we point out that the small excess
in the diphoton data obtained in a different experiment
\cite{ZPC69p561} also satisfies the here proposed distribution,
albeit with much lower statistics.

A further important ingredient for the reconstruction
of the present event distribution is the experimental resolution.
Given the $\eta$ signal in the COMPASS \cite{ARXIV11086191} data,
which is much broader than that of $\pi^{0}$,
and the fact that $\pi^{0}$ and $\eta$ have decay widths
considerably smaller than the bin size of 0.5 MeV,
we may assume that the experimental resolution
is better at lower invariant mass.

\begin{figure}[htpb]
\begin{center}
\begin{tabular}{c}
\includegraphics[width=200pt]{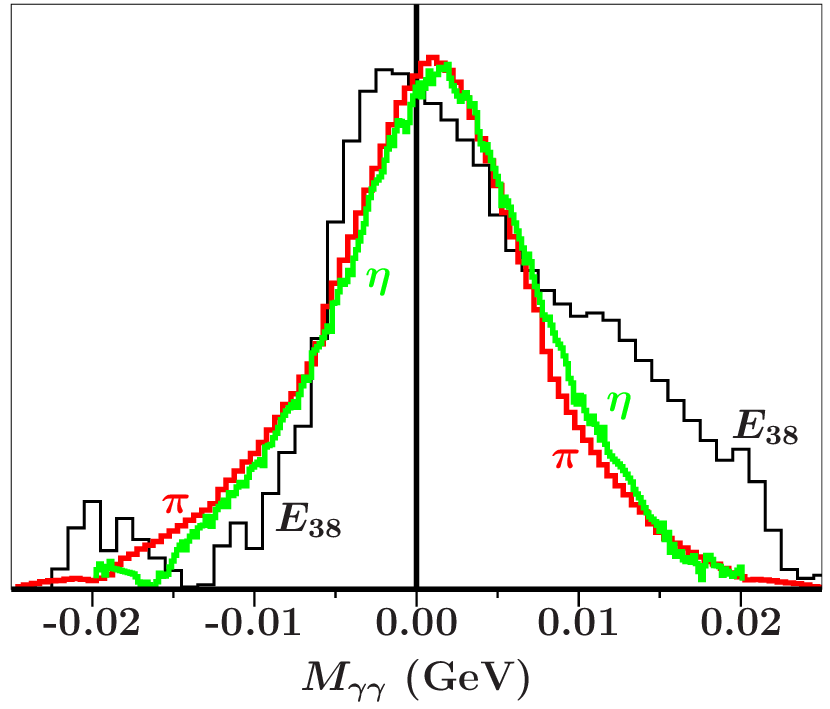}\\ [-20pt]
\end{tabular}
\end{center}
\caption{\small
The three enhancements observed by us in the $\gamma\gamma$
invariant mass distribution of Ref.~\cite{ARXIV11086191}:
$\pi^{0}$ (red),
$\eta$ (green),
and $E(38)$ (thin black curve).
The relative heights and widths of the three signals are in reality
different, viz.\ $\pi^{0}:\eta:E(38)\approx 1:0.06:0.03$
and $1:3.1:0.5$, respectively, but have been
adjusted here in order to compare the line shapes.
}
\label{width}
\end{figure}

In order to proceed, we compare in Fig.~\ref{width}
the three observed relevant structures
in the COMPASS diphoton data,
representing the $\pi^{0}$, $\eta$, and $E(38)$ bosons.
The three enhancements are scaled in width.
Also, the very different heights have been
adjusted in Fig.~\ref{width}, so as to allow comparing the line shapes.
The data spreading must be entirely due to the experimental resolution.
Furthermore, we assume that the $E(38)$ boson is reasonably stable
and so should have a spreading comparable to the other
two enhancements.
This seems to be confirmed by Fig.~\ref{width}, except
for the long $E(38)$ tail at higher invariant mass.

\begin{figure}[htpb]
\begin{center}
\begin{tabular}{c}
\includegraphics[width=230pt]{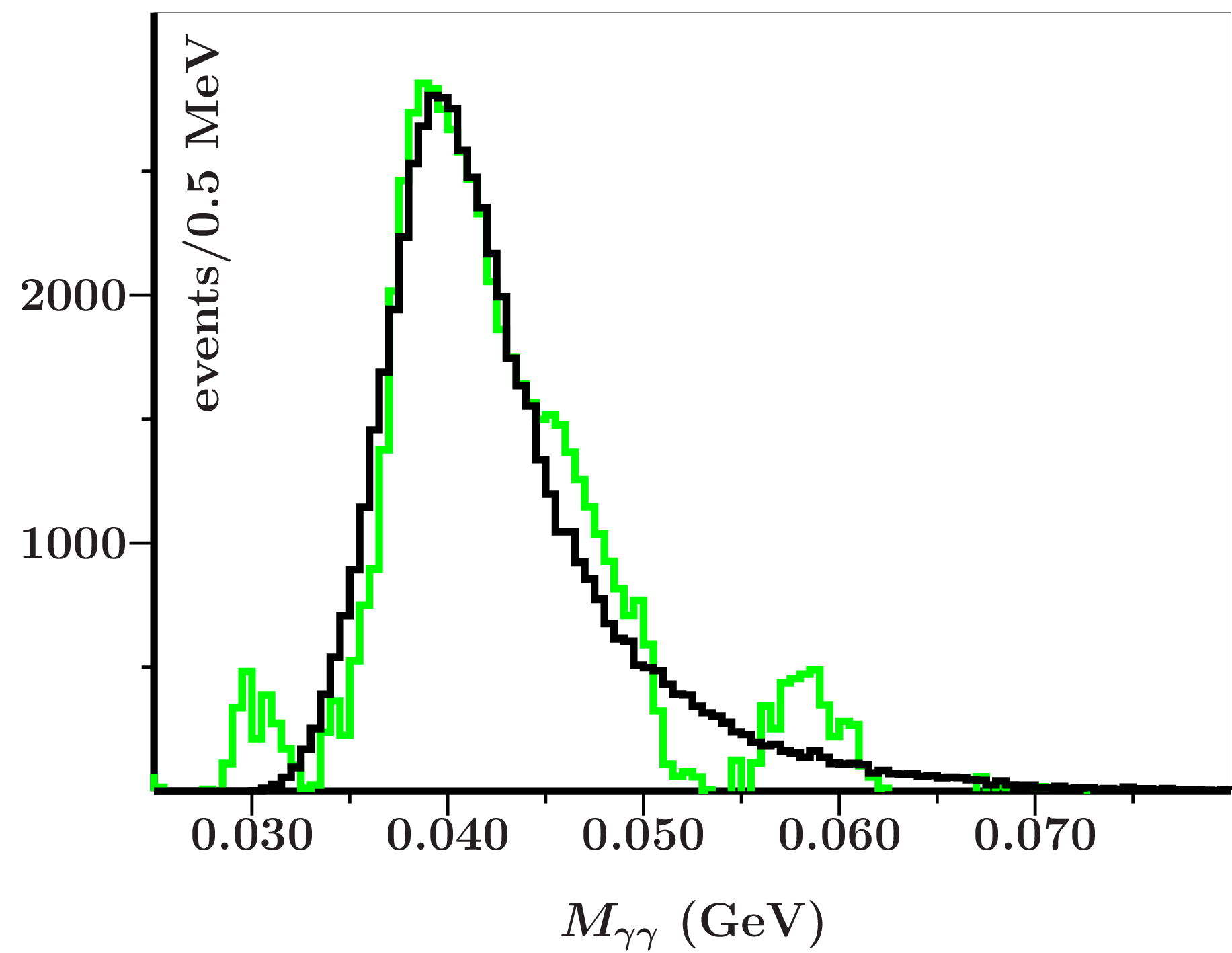}\\ [-20pt]
\end{tabular}
\end{center}
\caption{\small
Invariant two-photon mass distribution of Ref.~\cite{ARXIV11086191}
(green),
compared to model result (black).
}
\label{mccompass}
\end{figure}

From the representation of the enhancements in Fig.~\ref{width},
we infer an experimental resolution
of about 5.2 MeV in the 40 MeV region.
The convolution of the two Gaussians,
one for the spreading in the masses of the light bosons
and another for the experimental resolution around 40 MeV,
for 56,000 generated Monte Carlo (MC) events is show in
Fig.~\ref{mccompass}.

The here followed procedure is extremely simple,
but it reproduces the experimental result quite well
(see Fig.~\ref{mccompass}).
In particular, it describes how the peak shifts
to a little bit higher value than the expected 38 MeV,
and furthermore
reproduces the observed tail for larger invariant masses.
The central mass of the light boson allows for variations
of a few tenths of MeV, from which we may estimate its mass
to lie between 38.0 and 38.4 MeV.

We may thus conclude, with very simple assumptions, that
the event distribution of COMPASS near 40 MeV may
be understood by the production of a light boson
with a mass of 38.2$\pm$0.2 MeV.

The DELPHI Collaboration performed an analysis \cite{ZPC69p561}
of inclusive $\pi^{0}$ production in $Z^{0}$ decays.
The diphoton decays of $\pi^{0}$ were reconstructed by using pairs
of combinations of converted photons
as well as HPC photons, which are photons that were reconstructed
in the barrel electromagnetic calorimeter of the DELPHI detector.
In particular, reasonably high-statistics diphoton data
were shown by the DELPHI Collaboration
in Fig.~3b (1 converted photon and 1 HPC photon)
of Ref.~\cite{ZPC69p561},
which figure also contains a rather precise background fit.

\begin{figure}[htpb]
\begin{center}
\begin{tabular}{c}
\includegraphics[width=230pt]{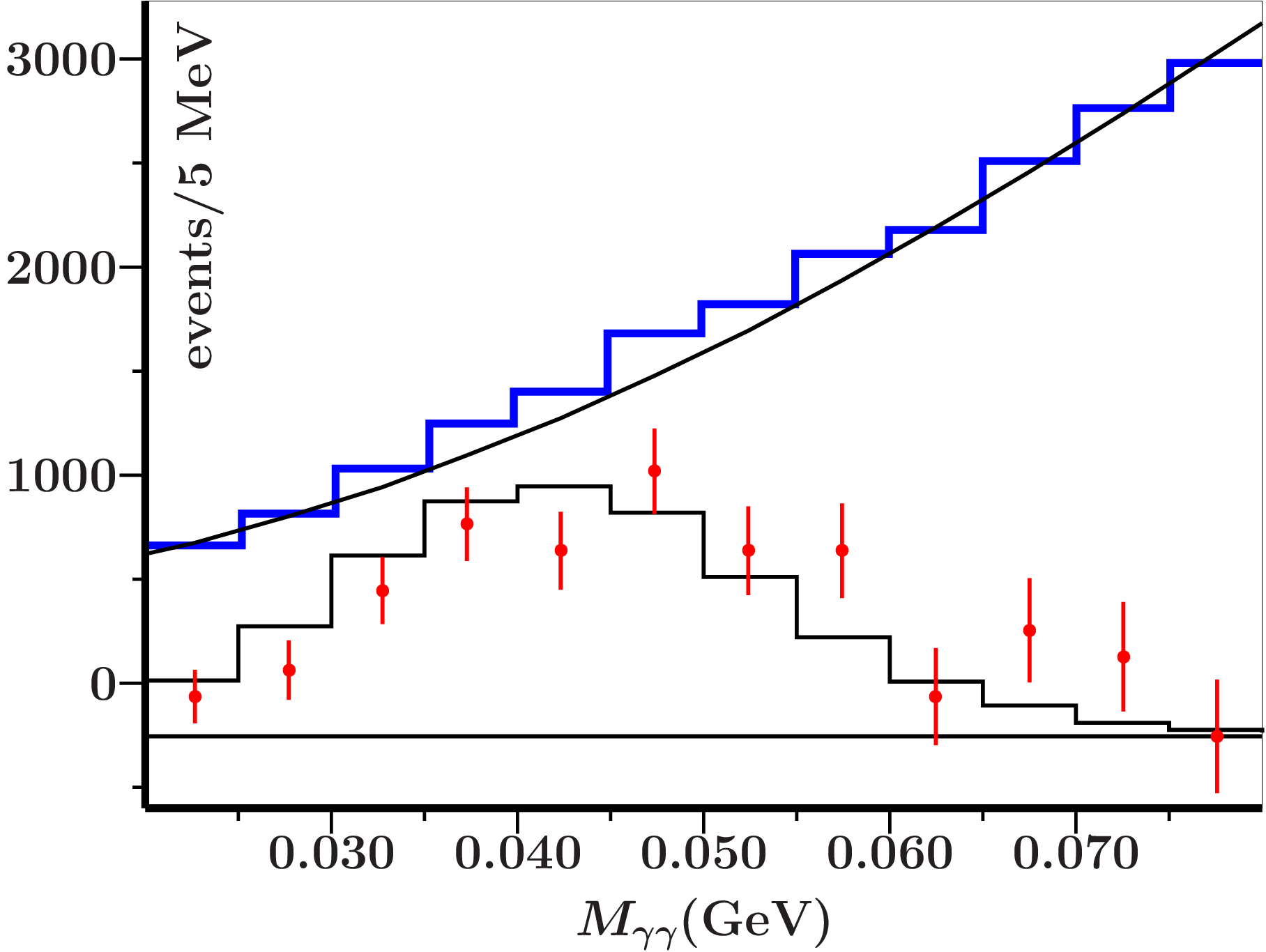}\\ [-20pt]
\end{tabular}
\end{center}
\caption{\small
Invariant two-photon mass distribution (blue)
and the background curve (black)
of Ref.~\cite{ZPC69p561}.
Bottom: difference between
the data and the DELPHI background curve,
multiplied by 5.
Thus, 1000 at the vertical axis represents 200 for the
dots and error bars (red).
The model result for 1400 MC data, represented by the lower
histogram, has been shifted downwards by 51 events/5 MeV,
in order to compensate for a mismatch between the background curve
and the data at lower invariant mass.
}
\label{mcdelphi}
\end{figure}

We apply a similar strategy to the residual signal
of the DELPHI data as we did above to the COMPASS data
of Ref.~\cite{ARXIV11086191}.
The result of the DELPHI data \cite{ZPC69p561},
with an assumed resolution of 10 MeV,
is depicted in Fig.~\ref{mcdelphi}.
So it seems to us that our assumption on the existence
of a scalar boson with a mass of about 38.2 MeV,
is very plausible,
since the setup and experimental conditions of the DELPHI experiment
are very different from those of COMPASS.

\section{Simulation of diphoton data}

\begin{figure}[htpb]
\begin{center}
\begin{tabular}{c}
\includegraphics[width=230pt]{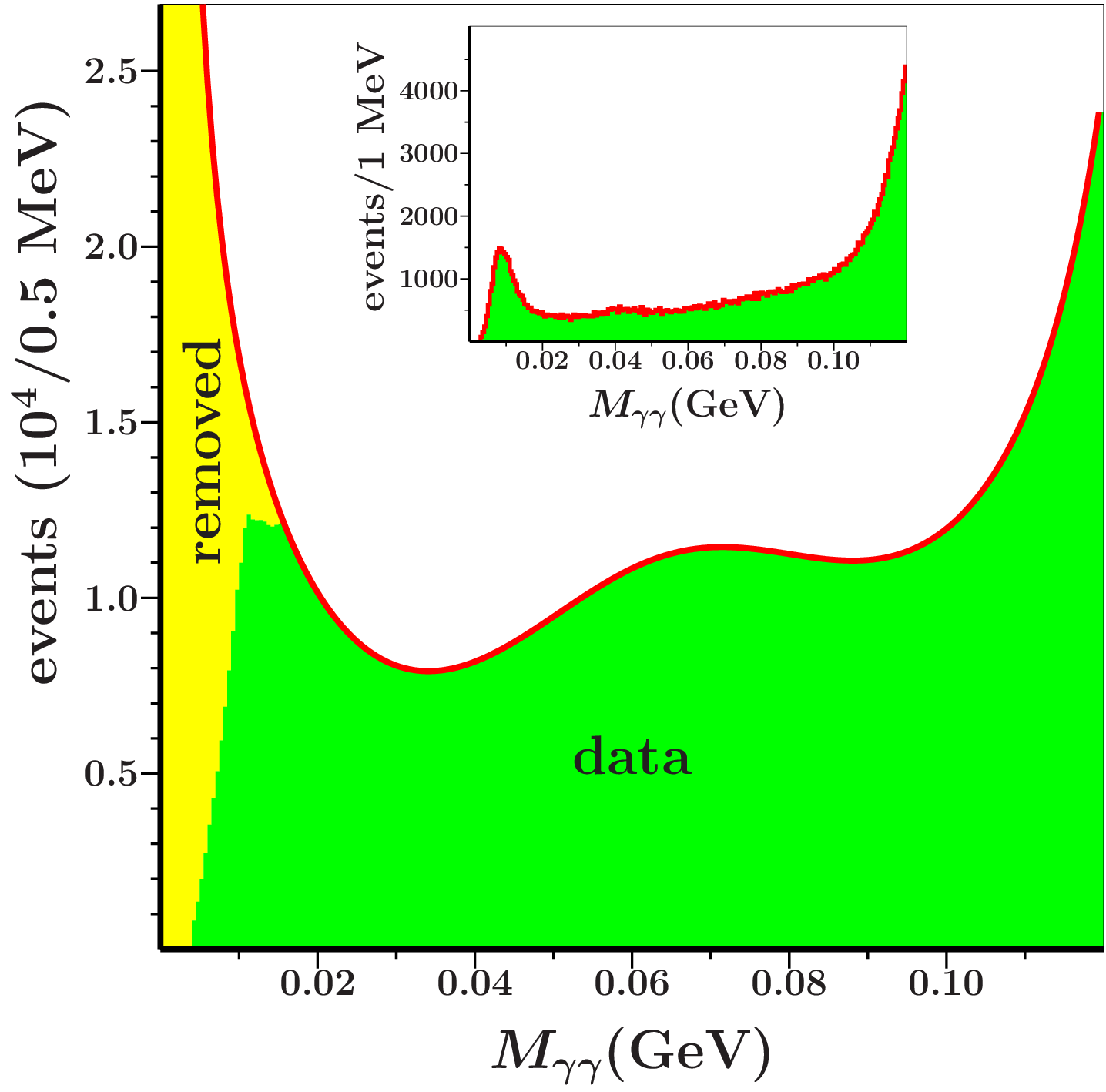}\\ [-20pt]
\end{tabular}
\end{center}
\caption{\small
Invariant two-photon mass distributions below the nominal
$\pi^0$ mass. The red curve in principle indicates the general aspect
of such a  distribution.
Data removed by the trigger system are represented by the yellow area.
So the green area in principle represents the final data.
The resulting low-mass peak of the main figure
coincides with the $\pi^{-}p$ data of Ref.~\cite{ARXIV11086191},
whereas, in the inset, we display the low-mass peak
for the $pp$ data of Ref.~\cite{ARXIV11090272}.
}
\label{ideal}
\end{figure}

In principle, the diphoton invariant-mass distribution
below the nominal $\pi^0$ mass has a structure as represented
by the red curve in Fig.~\ref{ideal}.
However, the two electromagnetic calorimeters,
ECAL1 at about 6 meters from the target
and ECAL2 at about 30 meters downstream,
do not accept low-energy photons.
This results in the non-observation
of the enhancement at zero invariant diphoton mass.
Hence, at very low masses no events are observed,
since they are removed by the trigger system
of the EM calorimeters, as indicated by the yellow area
in Fig.~\ref{ideal}.
It has the effect that at low mass a peak shows up in the data.

In Fig.~\ref{ideal} we depict the low-mass peak
for the $\pi^{-}p$ data of Ref.~\cite{ARXIV11086191},
while in the inset of Fig.~\ref{ideal},
we show the low-mass peak
for the $pp$ data of Ref.~\cite{ARXIV11090272}.
Note that these low-mass peaks are very different
for different experiments,
namely the one in the $pp$ data of Ref.~\cite{ARXIV11090272}
is roughly 3.8 times larger than the signal at 25 MeV,
whereas the one in the $\pi^{-}p$ data
of Ref.~\cite{ARXIV11086191} is only about 1.4 times higher
than the signal at 25 MeV.

Now, the data selection system does of course
also influence the aspect of the data for higher diphoton masses.
It may even result in some structures which resemble resonances.
Moreover, several physical processes in the experimental setup,
not related to the purpose of the experiment,
may result in further structures.

The authors of Ref.~\cite{ARXIV12042349} gave the following list
of mechanisms that may result in structures in their data.
\begin{itemize}
\item
Secondary $\pi^0$ mesons produced in the detector material
downstream of the target lead to diphoton masses which are
below the nominal $\pi^0$ mass when reconstructed
assuming a target vertex.
\item
Material concentrated in detector groups leads to peak-like structures.
\item
Secondary $e^+e^-$ pairs from photon conversion
in the spectrometer material lead to low-mass structures.
\item
Cuts applied in the reconstruction software lead to
additional structure in the low-mass range.
\end{itemize}
Most of those processes occur, of course, in the EM calorimeter
ECAL2, because it is further downstream than ECAL1,
and therefore picks up more contamination due to unwanted processes.

These artefacts are reproduced in the MC simulation
of Ref.~\cite{ARXIV12042349}
for the reactions under study, using a complete description
of the spectrometer material and employing
the same reconstruction software as for the real data analysis.

\begin{figure}[htbp]
\begin{center}
\begin{tabular}{c}
\includegraphics[width=240pt]{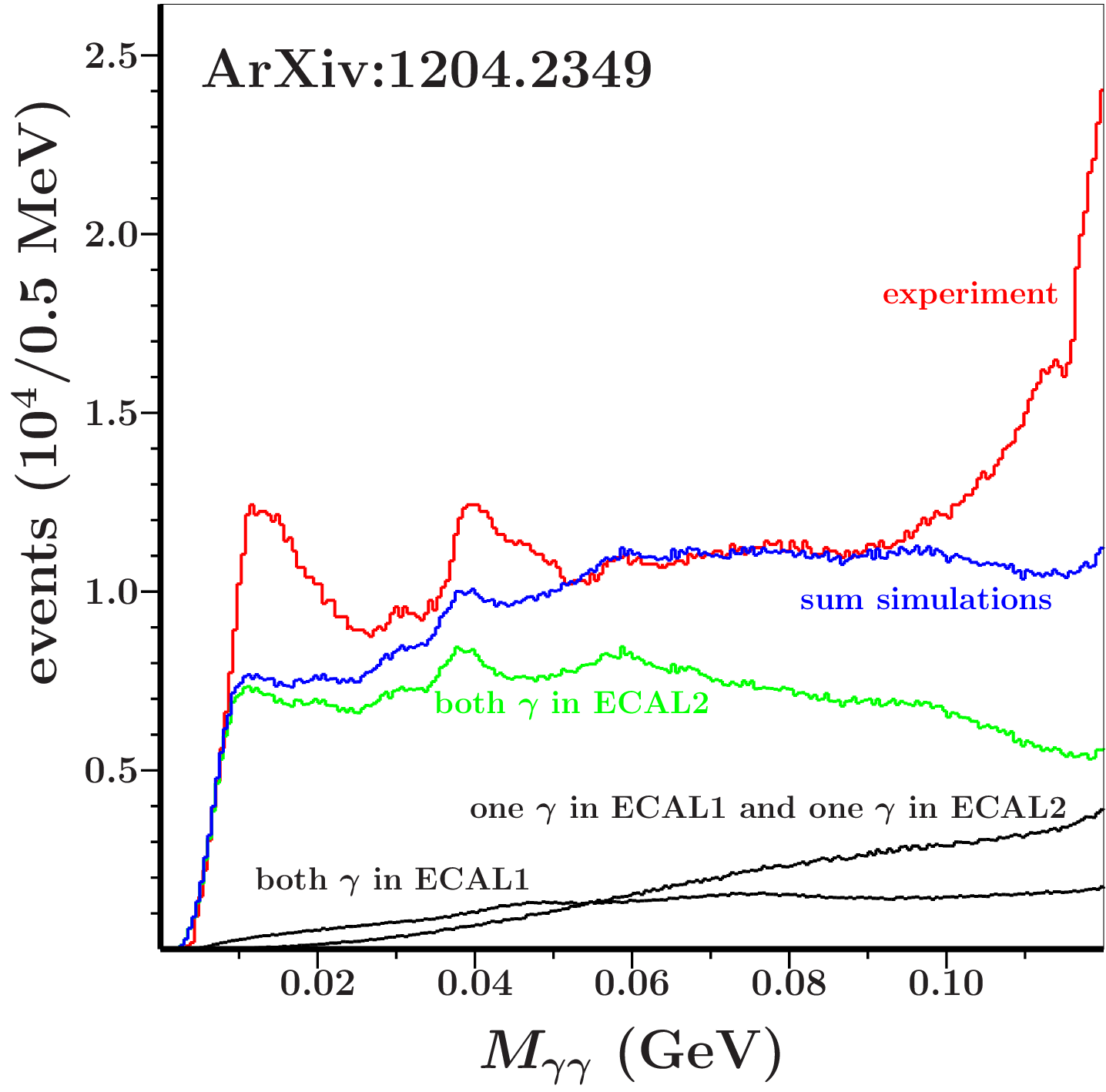}\\ [-10pt]
\end{tabular}
\end{center}
\caption{\small
The three diphoton MC distributions of Ref.~\cite{ARXIV12042349}
simulating low-energy $\gamma\gamma$ production, with
different weight factors.
Black curves: both $\gamma$s in calorimeter ECAL1 or one
in ECAL1 and one in ECAL2; green: both $\gamma$s in ECAL2.
Weight factors:
$\xrm{ECAL1}:\xrm{ECAL2}:\xrm{(ECAL1+ECAL2)}=0.48:1.00:0.86$.
Blue curve: sum of green and two black curves.
Red curve: the experimental data of Ref.~\cite{ARXIV11086191}.
}
\label{join}
\end{figure}

In Fig.~\ref{join} we reproduce the three independent simulations
reported in Ref.~\cite{ARXIV12042349}. We also add up these three
simulations, with relative weights
$\xrm{ECAL1}:\xrm{ECAL2}:\xrm{(ECAL1+ECAL2)}=0.48:1.00:0.86$
for both $\gamma$s in ECAL1,
both $\gamma$s in ECAL2, and one $\gamma$ in each of the two
electromagnetic calorimeters, respectively,
in order to have an idea how well they represent the true data
of Ref.~\cite{ARXIV11086191}. It should be noted, however, that
summing up three independent simulations is not the same as
performing an MC simulation for the complete experimental setup.
Therefore, our procedure is not totally reliable, though
it is the best we can do with the available information.

We observe from Fig.~\ref{join} that for some reason
the $\pi^{0}$ signal,
which in the experimental data sets out at about 50 MeV,
is not included in the simulations \cite{ARXIV12042349}.
Because of this omission, the simulated data for
one $\gamma$ observed in ECAL1
and another $\gamma$ in ECAL2 have to contribute more
than what one would expect from the COMPASS setup.

\begin{figure}[htbp]
\begin{center}
\begin{tabular}{c}
\includegraphics[width=240pt]{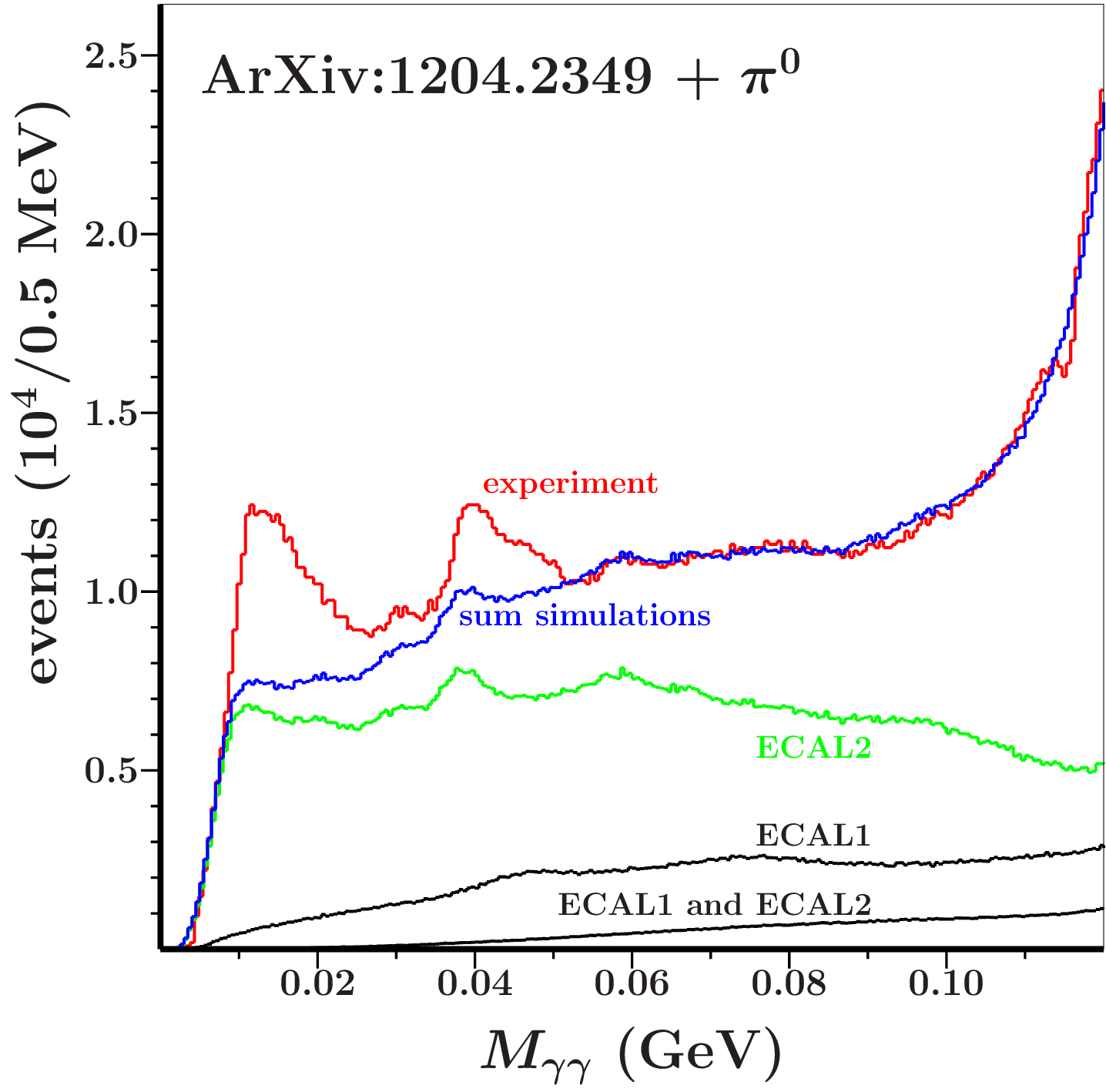}\\ [-10pt]
\end{tabular}
\end{center}
\caption{\small
As FIG.~\ref{join}, but now with weight factors
$\xrm{ECAL1}:\xrm{ECAL2}:\xrm{(ECAL1+ECAL2)}=0.80:0.93:0.25$,
and a simulation of the $\pi^{0}$ signal added in the blue
curve.
}
\label{joinpi}
\end{figure}

In Fig.~\ref{joinpi} we do include
the experimental signal of the $\pi^{0}$, with its mass
at 135 MeV and an average resolution of 12.7 MeV.
Once again, this is of course not the correct procedure.
But it is all we have at our disposal for a reconstruction
of the experimental data.
The simulations of Ref.~\cite{ARXIV12042349} are here taken in
proportions
$\xrm{ECAL1}:\xrm{ECAL2}:\xrm{(ECAL1+ECAL2)}=0.80:0.93:0.25$
of the reported MC data,
which seem more reasonable than the previous weights
employed in Fig.~\ref{join}.

At this point, we may conclude that experiment is well reproduced
for diphoton invariant masses larger than 50 MeV.
Nevertheless, we wonder how the MC distributions of
Ref.~\cite{ARXIV12042349} would work out for higher masses,
up to 1.0 GeV. However, for masses below 50 MeV,
the simulations of Ref.~\cite{ARXIV12042349}
do not at all agree with experiment.

In particular, the left-hand peak, near an invariant diphoton
mass of 10 MeV, is not reproduced at all.
This signal, which is almost entirely due
to data collected at the EM calorimeter ECAL2,
depends rather delicately on cuts in the data selection,
as we explained in the foregoing.
The result of Ref.~\cite{ARXIV12042349}
leaves us with the impression that,
in order to specifically enhance selected artefacts from ECAL2
in the 40 MeV region, the applied cuts in the MC simulation
are exagerated with respect to those actually employed in experiment.
However, it is also possible that the artefacts produced
in the COMPASS setup are just not well studied yet for small diphoton
masses.

\begin{figure}[htbp]
\begin{center}
\begin{tabular}{c}
\includegraphics[width=240pt]{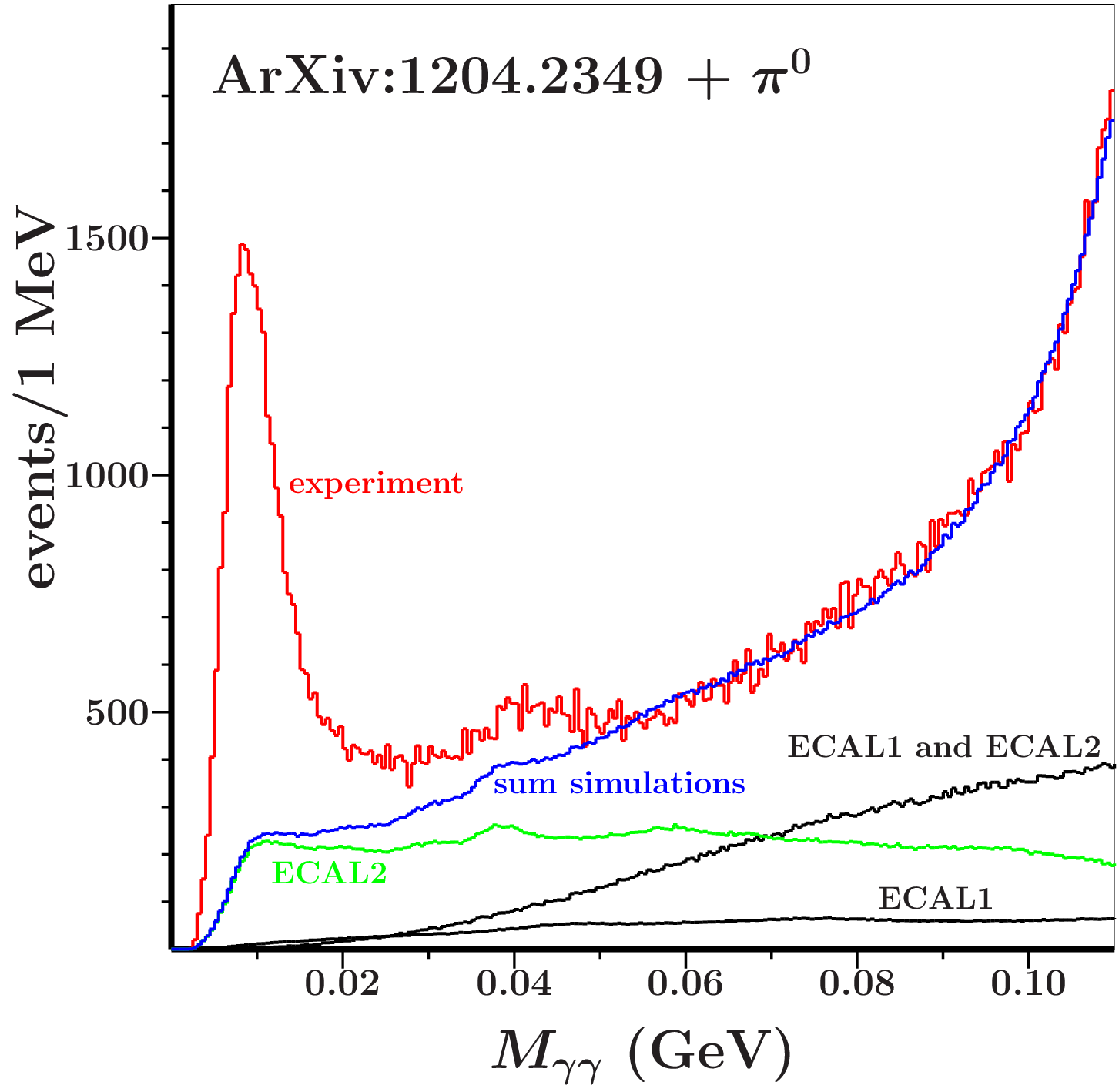}\\ [-10pt]
\end{tabular}
\end{center}
\caption{\small
Similar to FIG.~\ref{joinpi}, but now compared
to the actual experimental data of Ref.~\cite{ARXIV11090272}.
}
\label{ppjoinpi}
\end{figure}

The MC simulations shown in Ref.~\cite{ARXIV12042349}
were actually produced for the process studied in Ref.~\cite{ARXIV11090272}.
However, in that study the background-to-signal ratio
for the enhancement near 40 MeV is the same
as for the process considered in Ref.~\cite{ARXIV11086191}.
For completeness, we include in Fig.~\ref{ppjoinpi}
a comparison of the MC simulations of Ref.~\cite{ARXIV12042349}
and the actual experimental data of Ref.~\cite{ARXIV11090272}.
We find that also in this case the experimental data
below 50 MeV are not at all well described by the MC simulations
of Ref.~\cite{ARXIV12042349}.
The discrepancy between data and simulation at low diphoton masses
is even more serious.

In conclusion, we welcome the efforts of the authors
of Ref.~\cite{ARXIV12042349} to explain the observed enhancement
in the 35--51 MeV diphoton invariant-mass region
by conventional methods.
However, with the present MC simulations,
the existence of a resonance-like structure
at about 38 MeV cannot be excluded at all.
We suggest to include in future simulations the possibility
of $E(38)$ desintegration in the EM calorimeter ECAL2,
since we expect that this tentative light boson has rather stable
modes, which could easily survive the 30 meters
that separate ECAL2 from the target.

\section{Future}

Although the question whether there exists a (scalar) boson
with a mass of about 38 MeV does not depend exclusively
on the existence of a resonance-like structure
in the experimental data of Ref.~\cite{ARXIV11086191},
it is at present the clearest signal we have found in many
experiments. Diphoton data for the mass interval 10--100 MeV
are very rare and usually with low statistics.
Therefore, it is very important that the present issue be
settled, which requires a profound understanding
of all possible sources of artefacts.

\newcommand{\pubprt}[4]{#1 {\bf #2}, #3 (#4)}
\newcommand{\ertbid}[4]{[Erratum-ibid.~#1 {\bf #2}, #3 (#4)]}
\def\AIPCP{AIP Conf.\ Proc.}
\def\AdP{Annalen der Physik}
\def\AP{Ann.\ Phys.}
\def\CNPC{Chin.\ Phys.\ C}
\def\JMP{J.\ Math.\ Phys.}
\def\LNP{Lect.\ Notes Phys.}
\def\NCA{Nuovo Cim.\ A}
\def\PLB{Phys.\ Lett.\ B}
\def\PRD{Phys.\ Rev.\ D}
\def\PRL{Phys.\ Rev.\ Lett.}
\def\ZPC{Z.\ Phys.\ C}

\end{document}